\documentstyle[12pt]{article}

\begin{document}
\setbox1 = \hbox{{{$\lambda$}}}
\def\R{{{$\lambda$}}\hskip-\wd1{\raise2pt\hbox{$-$}}\hskip\wd1}
\def\lambar{\hbox{{\R}}}
\begin{titlepage}
\hoffset=.5truecm
\voffset=-2truecm

\centering

\null
\vskip 1truecm
{\huge \bf DIPARTAMENTO DI FISICA "G.Galilei"\\}
{\huge \bf ISTITUTO DI FISICA NUCLEARE \\}
{\bf Sezione di PADOVA \\
\bf ITALY}

\vskip 2truecm


{\LARGE \bf Strangeness Production in pp,pA,AA}  \\
\vskip 0.3cm
\begin{center}
{\LARGE \bf               Interactions }\\
\end{center}
\vskip 0.3cm
{\LARGE \bf      at SPS Energies. HIJING Approach }\\
\vskip 1truecm
{\large \bf  WA94 Collaboration}\\
\medskip
{\large \bf A.Andrighetto,M.Morando,F.Pellegrini\\}
\medskip
{\large \bf R.A.Ricci,G.Segato \\}
\medskip
Dipartimento di Fisica"G.Galilei",INFN Sezione di Padova,Via Marzolo 8,
35131 Padova,{\bf Italy\\}
\medskip
{\large \bf V.Topor Pop\\}
\medskip
{\normalsize Institute of Atomic Physics,P.O.Box MG-6,Bucharest},
{\bf Romania \\ }
{\normalsize and \\}
Dipartamento di Fisica "G.Galilei",INFN Sezione di Padova,Via Marzolo 8,
35131 Padova,{\bf Italy\\}
\medskip
\begin{flushleft}
{\bf DFPD-94-NP-42   \hspace*{9cm} 8 June 1994}
\end{flushleft}
\begin{center}
{\it Work presented at the International Conference
"Physics of High Energy Heavy Ion Collisions ",
Vuosaari(Helsinki),FINLAND,17-22 June 1994}
\end{center}
\vskip 1truecm

\end{titlepage}

\hoffset = -1truecm
\voffset = -2truecm

\title{\bf Strangeness Production in pp,pA,AA\\
             Interactions at\\
        SPS Energies.HIJING Approach}
\author{
{\bf V.Topor Pop\thanks{On leave from absence from Institute
 for Space Sciences,P.O.Box MG - 6,Bucharest,
{\bf Romania} E-MAIL TOPOR@ROIFA.BITNET(EARN);TOPOR@PADOVA.INFN.IT},
A.Andrighetto,M.Morando,F.Pellegrini,}\\
{\bf  R.A.Ricci,G.Segato}\\
\normalsize Dipartamento di Fisica "G.Galilei",Via Marzolo 8 -35131,
Padova \\
and INFN Sezione di Padova,
{\bf Italy}\\
}
\date{8 June  1994}

\maketitle

\begin{abstract}

In this report we have made a systematic study of strangeness
production in proton-proton(pp),proton-nucleus(pA) and nucleus-
 nucleus(AA) collisions at CERN Super Proton Synchroton energies,
using$\,\,\, HIJING\,\,\, MONTE \,\,\,CARLO \,\,\,MODEL $ \\
(version $HIJ.01$).

Numerical results for mean multiplicities of neutral strange
particles ,as well as their ratios to negatives hadrons($<h^{-}>$)
for p-p,nucleon-nucleon(N-N),\,\,p-S,\,\,p-Ag,\,\,p-Au('min.
bias')collisions and p-Au,\,\,S-S,\,\,S-Ag,\,\,S-Au
('central')collisions are compared to experimental data
 available from  CERN experiments and also with recent
 theoretical estimations given by others models($\,\,QGSM,
 \,\,RQMD,\,\,DPM,\,\,$ \\
 $\,\,VENUS$ and $FRITIOF$).

Neutral strange particle abundances are quite well described
for p-p,N-N and p-A interactions ,but are underpredicted by a
factor of two in A-A interactions for $\Lambda,\bar{\Lambda},
 K^{0}_{S}$ in symmetric collisions(S-S,\,\,Pb-Pb)and for
$\Lambda,\bar{\Lambda}\,\,$in asymmetric ones(S-Ag,\,\,S-Au,\,\,S-W).

The ratios antistrange/strange are well described for
$\frac{\bar{\Lambda}}{\Lambda}$ but the ratios for
 multistrange particles ($\Xi ,\Omega\,\,$)
 and their antiparticle, could not be predicted
by the model.A detailed analysis in limited rapidity range
is required in order to draw some definite conclusion.

A qualitative prediction  for rapidity, transverse kinetic
 energy and transverse momenta normalized distributions
are performed at 200 GeV/Nucleon in p-S,S-S,S-Ag and S-Au
 collisions in comparison with recent experimental data.
HIJING model predictions for coming experiments at CERN
for S-Au, S-W and Pb-Pb interactions are given.
The theoretical calculations are estimated in a full phase space.

The factor of two can not be explained by considering very central
events ($b_{max}=0.0$) and taking into account  minijets.
New ideas are necessary in order to  improve the comparison
 with experiment for nucleus - nucleus interactions.
\end{abstract}

\section{Introduction}

Searching for possible exotic states of nuclear matter is one of the
currently most interesting and stimulating subjects in high energy
nuclear physics (see references [1--10])
To create exotic matter such as the quark-gluon plasma (QGP)
[1--4,5]strange matter
\cite{kn:pro4},\cite{kn:raf1},multi$\,\,\Delta$-matter,pion
condensate,kaon condensate \cite{kn:pro4},\cite{kn:bono} it is
essential to generate extreme states of nuclear matter regarding
density, temperature or energy density \cite{kn:raf1}.
Such extreme states are considered to be produced most probably in
high energy heavy ion reactions[10--14]
or in the annihilation of energetic antiprotons($\bar{p} $)on a nucleus
\cite{kn:ahmad}

The possibility suggested by Witten \cite{kn:witt} that strange
quark-matter might be the lowest energy state of large baryon
number,would have many consequences in physics and astrophysics
and also cosmology relevance [16--20].
Small drops of strange matter(strangelets) may be natural results
of the evaporation of a QGP formed in heavy ion collisions.

Enhanced strangeness production \cite{kn:raf1} in ultrarelativistic
heavy ion collisions (URHIC) is one of the most widely discussed
signature for creating QGP \cite{kn:quer1},\cite{kn:quer2}.

Several experiments have been performed to investigate strangeness
production in reactions involving relativistic nuclei,particularly
at Super Proton Synchrotron (SPS) at CERN : NA35 [22--32],
NA36 [33--37],WA85,WA94[38--45],[128--130],NA44 [132--133]
 ,HELLIOS Collab.[46--50],
 and at the Alternating Gradient Synchrotron(AGS) at Brook-\\
haven E802 [51--55], E810 \cite{kn:1e810},\cite{kn:e810long} and
E814 \cite{kn:e814stac},[134].

These experiments have measured also the yield of several
particle species containing one or two strange quarks ($K^{+}$,
$K^{0}_{s}$,$\Lambda$,$\Xi^{-}$ and their antiparticles) or three
strange quarks($\Omega^{-}$).
In all the cases their production relative to pions
or negative hadrons and
some other ratios are larger in nucleus - nucleus collisions
compared to proton - proton .

Now experiments with truly heavy ion projectile have just
started(with a gold beam at BNL)\cite{kn:dieb93},\cite{kn:nam94}
 or are under way
($Pb(170 AGeV)+Pb$ at CERN)\cite{kn:quer2}.

The soft physics data have stimulated a great deal of theoretical
activity and in the present introduction it is impossible to give more
than a superficial overview of the main types of the Monte Carlo models
used.Many models for URHIC have been developed:Dual Partons Models(DPM)
[59--66],\cite{kn:dpm94}
Quark Gluon String Models(QGSM)[67--69],
VENUS models for Very Energetic NUclear Scattering [70--75]
 FRITIOF model
\cite{kn:ander},\cite{kn:nils},ATTILA \cite{kn:Gyu1} model,
relativistic quantum molecular dynamics
(RQMD) [79--83][135--138],Parton String Model(PSM)[139],
 HIJET model\cite{kn:fol94}.

 An excelent review and detailed comparision of the models
 was done by Werner \cite{kn:wer7}.

 On the other hand ,no unconventional explanation has so far put
 forward for the observed increase of antyhyperons
 ($\bar{\Lambda}$,$\Xi^{-}$).
 Actually ,it is the old Pomeron exchange picture that is revitalized
 in some models with a new ingredient that is the Pomeron is made of
 two colored strings.Thus the problem of interaction of strings
 becomes important.This interaction may give rice to the fusion of
 strings[85--88][137,138]
 making a new string with more color charge.These new
 objects could explain the strangeness enhancement \cite{kn:merino1},
 \cite{kn:merino2},[137,138]especially $\bar{\Lambda}$ enhancement.
 Such models can be seen as an interplay between
 independent string fragmentation and QGP formation.One version of
 VENUS generator also include such interaction \cite{kn:wer4},
 \cite{kn:merino2}.

 We note also that an interpretation based on a particular dependence
 of the cross sections for URHIC was taken into account \cite{kn:sal93}.

In this report we have performed a systematic study of strange particle
production in proton--proton($pp$),proton--nucleus($pA$) and nucleus--
nucleus ($AA$) interactions at SPS--CERN energies using HIJING(Heavy
Ion Jet Interacting Generator) Monte Carlo model [91--97]
 developed at Berkeley.

The formulation of HIJING is guided by the LUND FRITIOF
(\cite{kn:ander}--\cite{kn:nils}) and DPM \cite{kn:ranft88}
models phenomenology for soft nucleus - nucleus
reactions at intermediate energies $\sqrt{s} \le 20\,\, GeV/N$ and the
successfull implementation of perturbative QCD (PQCD) processes in
the PYTHIA model \cite{kn:pyt1}--\cite{kn:pyt2}
for hadronic interactions.

HIJING incorporate the PQCD to multiple jet processes and
the nuclear effects such as parton shadowing and jet quenching.
HIJING provide a link between the physics at intermediate CERN--SPS
energies and the highest collider energies at Relativistic Heavy Ion
Collider(RHIC)($\sqrt{s}\geq 200\,\, GeV/N$ and the CERN Large Hadron
Collider (LHC) ($\sqrt{s} = 6\,\, TeV/N$).

Using HIJING Monte Carlo model \cite{kn:wang2},\cite{kn:wang3} a
detailed discussion and comparison with a wide variety of data in
$pp$,$pA$ and $AA$ collisions was reported.However no study
 on the strangeness production at SPS-CERN energies in
$pp$,$pA$ and $AA$ data exist within this approach .

In this paper we present some theoretical calculations for mean
multiplicities of strange particle production in the full rapidity
range as well as their ratios to negative hadron
predicted by HIJING model(HIJ.01 version -see section
2 and 3 for explanation of this version)for $pp$ ,nucleon-nucleon
$N-N$,'minimum bias'collisions
$pS,pAg,pAu$ and\\
'central collisions'$\,\,pAu,SS,SAg,SAu,SW,PbPb\,\,$
 at the energy of $200\,\, AGeV$.

 Also a qualitative prediction for normalised distributions
 for rapidity,transverse kinetic energy and transverse
 momenta are done at 200 GeV/Nucleon in $pp,pS,SS,SAg$ and
 SAu collisions and the results are compared with recent
 experimental data from NA 35 Collaboration[140].We note that
 theoretical calculations are done for full phase space.
 For giving an idea of abundances and spectra in experiments
planned in the near future at CERN ,we give some predictions
for SW and PbPb interactions.

We note that the version HIJ.01 of the model includes neither the
formation of QGP nor some collective dynamics on either the hadronic
or string level.Although the existing trends of hadron distributions
are quite satisfactorily approximated by the HIJING model
\cite{kn:wang2},\cite{kn:wang3}and also neutral strange production
in $pp$ and $pA$ interactions are quite well described,
this version is getting less reliable
for strange particle produced in heavy colliding systems.
The  same result is stressed out and in other models
QGSM\cite{kn:ame1},\cite{kn:ame2},RQMD\cite{kn:sorge2},
FRITIOF\cite{kn:6na35},PSM\cite{kn:ame94}.

A brief description of the HIJING Monte Carlo model and theoretical
background are given in Section 2.Detailed numerical results of
this version HIJ.01 for $\,\,pp,pA\,\,$ and $\,\,AA\,\,$
reactions especially at
CERN-SPS energies$\,\,\sqrt{s}\simeq 20\,\, GeV/N$ for
strangeness production
are compared to experimental data and other models predictions like
QGSM \cite{kn:ame1},\cite{kn:ame2},
DPM \cite{kn:ranft3},[141],142],VENUS \cite{kn:wer7}
,FRITIOF \cite{kn:6na35},
and RQMD \cite{kn:sorge2},[140] in Section 3. Section 4 concludes with
a summary and discussion of future applications.

\section{The Monte Carlo HIJING model}

A detailed discussion of the HIJING Monte Carlo model was reported
in references \cite{kn:wang1},\cite{kn:wang2},\cite{kn:wang4}.
The formulation of HIJING was guided by the LUND-FRITIOF and DUAL
PARTON MODEL phenomenology for soft nucleus-nucleus reactions at
intermediate energies ($\sqrt{s}<20\,\, GeV$)and implementation
of perturbative QCD(PQCD) processes in the PHYTHIA MODEL for hadronic
interactions.Comparison between some MONTE CARLO GENERATORS are
given in reference \cite{kn:wer7}

All PQCD models employ initial and final state radiation in an
iterative fashion\,\,:\,\,a parton with a squared mass $Q^{2}$ radiates
a parton and leaves a remainder parton with reduced $Q^{2}$,the
latter one radiating again and so on .The hadronization finally
is done by ussing string model(PYTHIA).A link to the soft
semihard models is provided by PYTHIA ,where it was realized that
the perturbatively calculated cross section is an inclusive one
and may well exceed the total cross section meaning multiple
scattering.Therefore the cross sections was eikonalized:for a
given impact parameter multiple scattering occurs acording to a
Poissonian distribution,with the average number of scatterings,
being $b$--dependent,proportional to some profile function
$T(b)$.HIJING is based on a particular model of high energy
pp inelastic collisions.

We give in this section brief review of the aspect relevants to
hadronic interaction:
\begin{enumerate}
\item  The models includes multiple minijet production with initial
and final state radiation along the lines of the PYTHIA model
\cite{kn:pyt1},\cite{kn:pyt2} and with cross sections calculated
within the eikonal formalism.
\item Soft beam jets are modeled by quark-diquark strings with gluon
kinks along the lines of the DPM and FRITIOF models.Multiple low
$p_{T}$ exchanges among the end point constituents are included.
\item Exact diffuse nuclear geometry is used to calculate the impact
parameter dependence of the number of inelastic processes
\cite{kn:Gyu1}.
\item An impact parameter dependent parton structure function is
introduced to study the sensitivity of observable to nuclear
shadowing,especially of the gluon structure functions.
\item A model for jet quenching is included to enable the study of
the A-dependence of moderate and high $p_{T}$ observable on an
assumed energy loss of partons traversing the produced dense matter.
\item HIJING does not incorporate the mechanism for final state
interactions among low $p_{T}$ produced particles.
\end{enumerate}

The rate of multiple minijet production in HIJING is constrained by
the cross sections in nucleon-nucleon collision .Within an eikonal
formalism \cite{kn:hwa1} the total elastic cross sections
$\sigma_{el}$,total inelastic cross sections $\sigma_{in}$ and
total cross sections $\sigma_{tot}$ can be expressed as:
\vskip 0.3cm
\begin{equation}
\sigma_{el}=\pi\int_{0}^{\infty}\:db^{2}(1-\exp(-\chi(b,s)))^{2}
\label{e1}
\end{equation}
\vskip 0.3cm
\begin{equation}
\sigma_{in}=\pi\int_{0}^{\infty}\:db^{2}(1-\exp(-2\,\chi(b,s)))
\label{e2}
\end{equation}
\vskip 0.3cm
\begin{equation}
\sigma_{tot}=2\,\pi\int_{0}^{\infty}\:db^{2}(1-\exp(-\chi(b,s)))
\label{e3}
\end{equation}
\vskip 0.3cm
Strong interactions involved in hadronic collisions can be
generally divided into two categories depending on the scale of
momentum transfer $q^{2}$ of the processes.
If $q^{2}< \Lambda_{QCD}^{2}$ the collisions are
nonperturbative and are considered {\it soft}\,\,and if
$q^{2}\gg \Lambda_{QCD}^{2}$ the subprocesses on the parton
level are considered {\it hard} and can be calculated via PQCD
\cite{kn:wang3}.

In the limit that the real part of the scattering amplitude is
small and the eikonal function $\chi(b,s)$\,\,is real ,the factor
\vskip 0.3cm
\begin{equation}
g(b,s)=1-\exp(-2\,\chi(b,s))
\label{e4}
\end{equation}
\vskip 0.3cm
in terms of semiclassical probabilistic model can be interpreted as
{\sl the probability for an inelastic event of nucleon--nucleon
collisions at impact parameter $b$ ,which may be caused by hard,
semihard or soft parton interactions}.

To calculate the probability of multiple minijet a main dynamical
assumption is that they are independent.This holds as long as
their average number is not too large.The independence should
apply up to LHC energies \cite{kn:wang3}

When shadowing can be neglected,the probability of no jets and
$j$ independent jet production in an inelastic event at impact
parameter $b$,can be written as :
\vskip 0.3cm
\begin{equation}
g_{0}(b,s)=(1-\exp(-2\,\chi_{s}(b,s)))\exp(-2\,\chi_{h}(b,s))
\label{e5}
\end{equation}
\vskip 0.3cm
\begin{equation}
g_{j}(b,s)=\frac{\left[2\,\chi_{h}(b,s)\right]^{j}}{j!}\cdot
            exp(-2\,\chi_{h}(b,s))\,\,\,\,\,\,  j \geq 1
\label{e6}
\end{equation}
\vskip 0.3cm

where $\chi_{s}(b,s)$ --is the eikonal function for soft interaction,
$2\,\chi_{h}(b,s)$ --is the average number of hard parton interactions
at a given impact parameter,
$\exp(-2\,\chi_{s}(b,s))$ --is the probability for no soft interaction.
Summing eqs.(5) and (6) over all values of $j$ leads to :
\vskip 0.3cm
\begin{equation}
\sum_{j=0}^{\infty}\,\,g_{j}(b,s)=1-\exp(-2\,\chi_{s}(b,s)-2\,
\chi_{h}(b,s))
\label{e7}
\end{equation}
\medskip

Comparing with eq.(\ref{e4}) one has :
\vskip 0.3cm
\begin{equation}
\chi(b,s)=\chi_{s}(b,s)+\chi_{h}(b,s)
\label{e8}
\end{equation}
\vskip 0.3cm
If we consider that the parton distribution function is factorizable
in longitudinal and transverse directions and the shadowing can be
neglected the average number of hard interaction $2\chi_{h}(b,s)$
at the impact parameter $b$ is given by :
\vskip 0.3cm
\begin{equation}
\chi_{h}(b,s)=\frac{1}{2}\,\,\sigma_{jet}(s)\,T_{N}(b,s)
\label{e9}
\end{equation}
\vskip 0.3cm
where $T_{N}(b,s)$ is the effective partonic overlap function of the
nucleons at impact parameter $b$.
\vskip 0.3cm
\begin{equation}
T_{N}(b,s)=\int d^{2}b'\rho(b')\rho(\left|b-b'\right|)
\label{e10}
\end{equation}
\vskip 0.3cm
with normalization
\vskip 0.3cm
\begin{equation}
\int\,\,d^{2}b\,T_{N}(b,s)=1
\label{e11}
\end{equation}
\vskip 0.3cm
and$\,\, \sigma_{jet}\,\,$is the is the PQCD cross section
of parton interaction
or jet production \cite{kn:wang2},\cite{kn:wang3}

If we note $\,\,\xi=b/b_{0}(s)\,\,$,where $\,\, b_{0}(s)\,\,$provide a
measure of the geometrical size of the nucleon
$$\pi b_{0}^{2}(s)=\sigma_{s}(s)/2$$
we get assuming the same geometrical distribution for both
soft and hard overlap functions
\vskip 0.3cm
\begin{equation}
\chi_{s}(\xi,s)\equiv \frac{\sigma_{s}}{2\sigma_{0}}\chi_{0}(\xi)
\label{e12}
\end{equation}
\begin{equation}
\chi_{h}(\xi,s)\equiv \frac{\sigma_{jet}}{2\sigma_{0}(s)}\chi_{0}(\xi)
\label{e13}
\end{equation}
\begin{equation}
\chi(\xi,s)\equiv \frac{1}{2\sigma_{0}}\left [{\sigma_{s}(s)
+\sigma_{jet}(s)}\right ]\chi_{0}(\xi)
\label{e14}
\end{equation}
\vskip 0.3cm
We note that$\,\, \chi(\xi,s)\,\,$ is a function not only of $\,\,\xi\,\,$
 but also of
$\,\,\sqrt{s} \,\,$because of the $\,\,\sqrt{s}\,\, $ dependence on
the jet cross
section $\,\,\sigma_{jet}(s)\,\,$.Geometrical scaling implies on the other
hand,that $\,\,\chi_{s}(\xi,s)=\chi_{0}(\xi)\,\,$ is only a function
 of$\,\,\,\xi$.
Thefore geometrical scaling is broken at high energies by the
introduction of the nonvanishing $\,\sigma_{jet}(s)\,$ of jet production.

Now we can rewrite the cross sections of nucleon - nucleon collisions
as :
\vskip 0.3cm
\begin{equation}
\sigma_{el}=\sigma_{0}(s)\int_{0}^{\infty}d\,\xi^{2}\left ( 1-exp(-\chi
(\xi,s)\right )^{2}
\label{e15}
\end{equation}
\begin{equation}
\sigma_{in}=\sigma_{0}(s)\int_{0}^{\infty}d\,\xi^{2}\left ( 1-exp(-2\,
\chi(\xi,s)\right))
\label{e16}
\end{equation}
\begin{equation}
\sigma_{tot}=2\,\sigma_{0}(s)\int_{0}^{\infty}d\,\xi^{2}\left (1-exp
(-\chi(\xi,s)\right))
\label{e17}
\end{equation}
\vskip 0.3cm
The calculation of these cross sections requires specifying
$\sigma_{s}(s)$ with a coresponding value of cut - off momenta
$P_{0}$ \cite{kn:wang4},\cite{kn:wang5}

In the energy range $10\,\, GeV<\sqrt{s}<70\,\, GeV$,where only soft
parton interactions are important cross sections
$\sigma_{s}(s)$ is fixed by the data on total cross sections
$\sigma_{tot}(s)$ directly.In and the above $Sp\bar{p}S$ energy range
$\sqrt{s}\geq 200\,\, GeV$  we fix $\sigma_{s}(s)$ at a value of
57 mb with $P_{0}=2 \,\,GeV/c$,in order to fit the data of the
cross sections.Between the two regions
$70\,\, GeV < \sqrt{s}<200\,\, GeV$,we simply use a smooth extrapolation
for $\sigma_{s}(s)$\cite{kn:wang5}.

In HIJING a nucleus-nucleus collisions is decomposed into binary
collisions involving in general excited or wounded nucleons.
Wounded nucleon are assumed to be $\,q-qq\,$ string like
configurations that decay on a slow time scale compared to the
collision time of the nuclei.In the FRITIOF scheme wounded nucleon
interactions follow the same excitation law as the original hadrons.
In the DPM scheme subsequent collisions essentially differ from
the first since they are assumed to involve sea partons instead
of valence ones.The HIJING model adopt a hybrid scheme,iterating
string-string collisions as in FRITIOF but utilizing DPM like
distributions.

By incorporating the successful multistring phenomenology for
low $p_{T}$ interactions at intermediate energies,HIJING provide
a link between the dominant nonperturbative fragmentation physics
at intermediate CERN-SPS energies and perturbative QCD physics
at the highest collider energies(RHIC;LHC)

\section{NUMERICAL RESULTS}
\subsection{STRANGENESS IN PROTON - PROTON INTERACTION}

Strange particle production provide some information about the
question of whether or not a quark-gluon plasma occurs in heavy
ion collisions.This is major motivation to first study the
production of strange particles in pp scattering.In addition,
strange particle give some insight into dynamics already for pp
collisions.

We run the program HIJING , with default parameters,mainly
 IHPR2(11)=1 ,which means choise of baryon production
model with diquark-antidiquark pair production allowed,initial
diquark treated as unit ; IHPR2(12)=1 ,decay of particle such as
$\pi^{0}$,$K_{s}^{0}$,$\Lambda$,$\Sigma$,$\Xi$,
$\Omega$ are allowed\,\,;\,\,IHPR2(17)=1
 - gaussian distribution of transverse
momentum of the sea quarks ;IHPR2(8)=0 - jet production turned
off for theoretical predictions in HIJING model- HIJ.01,
 and IHPR2(8)=10-the maximum
number of jet production per nucleon-nucleon interaction
for  for theoretical predictions
$\,\,HIJ.01^{(j)}\,\,$.

The energy dependence of average multiplicities for different particle
species,gives a first impression about the predictive power of the
model and provide an important check of model approach.

In Table 1 we give average multiplicities of particle at
$E_{lab}=200\,\, GeV\,\,$ in proton-proton interaction,
$E_{lab}$ being the laboratory energy.

The theoretical values $\,\,HIJ.01\,\,$ are obtained for
$\,\,10^{5}\,\,$ generated events and in a full phase space.
The values $\,\,HIJ.01^{(j)}\,\,$ are for very central $pp$
collisions($b_{max}=0.0\,\,$) and  with minijet production.
The experimental data are taken from reference
\cite{kn:11na35}.

The large difference between pions and kaons is due to the suppressed
strangeness production from string fragmentation.Pions($+$) and
kaons($+$) are  more frequent than the negative ones due to
charge conservation.We note that the multiplicities for neutral
strange particle $<\Lambda>,<\bar{\Lambda}>,<K_{s}^{0}>$ and
for antiproton
$<\bar{p}>\,\,$are quite well described in the limits of three
standard error deviations for
 $pp$ interactions at  $\,\,200\,\, GeV\,\,$ .However the values for
$<\bar{p}>$  and $<\bar{\Lambda}>$ are slightly overpredicted
by the model.

In order to give some comparision at ultrahigh energy and to test
the predictive power of generator  we perform
calculations at centre of mass energy
$\sqrt{s}=546\,\, GeV$($Sp\bar{p}S$-energies),
for $\,\,\bar{p}p\,\,$ interactions.

By using many pieces of data from the different collider experiments
(charge particle $K,\Lambda,\Xi$ and $\gamma$ from {\bf UA5} ,$p$ from
{\bf UA2}) ,the {\bf UA5} collaboration has attempt to piece together
a picture of the composition of a typical soft event at the
$Sp\bar{p}S$  \cite{kn:ward}.The measurements were made in various
different kinematic regions and have been extrapolated in the full
transverse momenta($p_{T}$) range and rapidity range for comparison
as described in reference \cite{kn:1ua5}.
The experimental data are compared with $HIJ.01^{(j)}$ results in Table 2.
The number of generated events was $\,\, 10^{5}\,\,$.

 It was stressed out \cite{kn:ward} that
 the data show a substantial excess of photons compared to the
 mean $\pi^{+}+\pi^{-}$.It was suggested like explanation the
 possible emission of gluon Cerenkov radiation in hadronic
 collision \cite{kn:drem}.At this energy the model $\,\,HIJ.01^{(j)}\,\,$
 seems to work fairlly well and the data are  well described.
 Our calculations ruled out such hypothesis.Taken into account
direct gamma production the agreement with the data is in the
limit of experimental errors.

 The experimental ratio $\frac{K^{+}}{\pi^{+}} =0.095 \pm 0.009$ are
 predicted also by$\,\, HIJ.01^{(j)}\,\,$ model (0.99).
 We note that a
 study of the ratios of invariant cross sections of kaons to that
 of pions as a function of transverse momenta in the central region
 was done by  Wang and Gyulassy \cite{kn:wang2},\cite{kn:wang3}.

 The behaviour of average multiplicities for different particles
 species as a function of the energy can be reproduced by
 HIJING model\cite{kn:wang3} and also by
 VENUS model(see reference \cite{kn:wer7}).

 Appropriate variable to describe single particle properties are
 the transverse momentum $p_{T}$ and the rapidity $y$
 \vskip 0.3cm
 \begin{equation}
 y=\frac{1}{2}ln \frac{E+p_{3}}{E-p_{3}}=ln\frac{E+p_{3}}{m_{T}}
 \label{e18}
 \end{equation}
 \vskip 0.3cm
 with $E,p_{3}$,and $m_{T}$ being energy,longitudinal momentum and
 transverse mass
 \vskip 0.3cm
 \begin{equation}
 m_{T}=\sqrt{m_{0}^{2}+p_{T}^{2}}
 \label{e19}
 \end{equation}
 \vskip 0.3cm
 with $m_{0}$ being particle mass.

 In figures 1a,b and 2a,b we show normalized rapidity (1a,2a)and
 transverse momentum distributions(1b,2b) for $\,\,\Lambda\,\,$'s
 (fig.1a,b) and for $\,\,K_{s}^{0}\,\,$'s for$ pp$ scattering
 at$ 200\,\, GeV$ .  The theoretical values
HIJ.01  are compared with experimental
 data taken from Jaeger et al.\cite{kn:jagl}
 HIJING results and data \cite{kn:jagl}
 are very similar ,although $\,\,HIJING\,\,$
  does not show some structures present in data.

We see the two distinct lambda peaks ,being due
 (in zeroth order)the fragmentation of the forward baryonic string
 and the backward baryonic string into leading baryons.Therefore
 lambda distributions are similar to proton distributions,just
 reduced by some factor and with the diffractive peak missing(see
 also reference \cite{kn:wer7}).

 The kaons are peaked at mid rapidity and are similar to the
 pions,though again less in magnitude.The distributions are
 narrower compared to proton's and pion's.

\subsection{STRANGENESS IN PROTON-NUCLEUS
      AND NUCLEUS-NUCLEUS INTERACTIONS}
\subsubsection{Multiplicities in $pA$ and $AA$ collisions}

 In the following we investigate strange particle production
 in HIJING model approach,for proton-nucleus and nucleus-
 nucleus collisions.
 We have studied the production of neutral strange particle
 $\Lambda,\bar{\Lambda},K_{s}^{0}$
 in comparison with recent experimental data from NA35
 CERN-experiments,for $pp,pA,AA$ interactions.
 We investigate the average multiplicities for negative hadrons
$<h^{-}>$,negative pions($\,\,<\pi^{-}>\,\,$
 and neutral strange particles ($<K_{s}^{0}>$,$<\Lambda>$,
 $<\bar{\Lambda}>$) in $\,\,pp,NN,pS,pAg,pAu\,\,$
'minimum bias collisions'and  $\,\,SS,SAg,SAu\,\,$
'central collisions' at  $200\,\, GeV/N\,\,$.
 The default parameters were used during the simulation and also
 option JET NO(IHPR2(8)=0).
 The number of Monte Carlo generated events was $\,\,10^{5}\,\,$
for $\,\,pp\,\,$ and  $\,\,pA\,\, $interactions ,$\,\, 5\cdot 10^{3}\,\,$
for$\,\,SS,SAg\,\,$ and $\,\,10^{3}\,\,$ for $\,\,SW,SAu\,\,$
and $\,\,PbPb\,\,$ collisions.

 The results are given in Tabel 3 in comparison with
 experimental data \cite{kn:na3594}.The corresponding
 data for $pp$ and $NN$ at 200 GeV \cite{kn:11na35} are also
 shown in Table 3.

 We remark that HIJING model describe quite well neutral
 strange particle multiplicities for $\,\,pp\,\,$ and
 $\,\,pA\,\,$ interactions ,but underestimate by a factor
 of two the values for nucleus-nucleus interactions for
 $<\Lambda>$,$<\bar{\Lambda}>$,$<K_{s}^{0}>$ in symmetric collisions
 $\,\,SS,PbPb\,\,$(expected values) and for $<\Lambda>$,
 $<\bar{\Lambda}>$ in assymmetric ones
 ($\,\,SAg,SAu,SW\,\,$).

 We note also that our theoretical calculations are done for
 $\,\,pA\,\,$ 'minimum bias' collisions
 and the experimental data are
 for the events with charged particle multiplicity greater than
 five ,which contain a significant fraction (about 90 \%) of the
 'minimum bias' events \cite{kn:na3594}.

 In order to get some idea about the discrepancies between
 theoretical calculations and experimental data we try
  to run the program in the limiting condition for
 very central events ($\,\,b_{max}=0.0\,\,$) and jet allowed.
 The results are given in Table 4 and are compared also
 with recent theoretical approach in RQMD \cite{kn:na3594},
 QGSM \cite{kn:ame2},\cite{kn:na3594} and DPM models
($\,\,DPM^{1}\,\,$ are from reference \cite{kn:mohr1},
 variant of DPM which include additionally
  $\,\,(qq)-(\bar{q}\bar{q})\,\,$ production
 from the sea into the chain formation process,
  $\,\,DPM^{2}\,\,$ are from reference \cite{kn:mohr2},
 variant of DPM which include chain fusion ,good candidate
 to explain the anomalous antihyperon production).

 In order to study the total production of strangeness in
 these collisions in a model independent way and to use
 all available experimental information NA35 Collaboration
 \cite{kn:na3594} has studied the ratio defined as :
\vskip 0.3cm
\begin{equation}
E_{S}=\frac{<\Lambda>+4<K_{s}^{0}>}{3<\pi^{-}>}
\label{e20}
\end{equation}

\vskip 0.3cm

 We estimate this ratio in HIJING approach for the above
 interactions and give the numerical results in Table 5
together with the mean multiplicities for negative pions.
We can conclude from this analysis that the factor of
two for $\,\,<K_{s}^{0}>\,\,$ which was reclaimed from
multiplicities analisis for symmetric interactions are
ruled out.Experimental error are to higher and new data
with higher  statistics are needed for draw a definite
conclusions.

Taken into account a factor of two in neutral strange
particle production we get a value $0.26-0.28$
independent of the  target nucleus mass number  for $AA$
interaction from $\,\,SS\,\,$ to $\,\,PbPb\,\,$.
Without factor of two the theoretical $\,\,E_{S}\,\,$
values for $\,\,pA\,\,$ and $\,\,AA\,\,$ are aproximatively equal
to the values for nucleon-nucleon interactions.
However the experimental values of $\,\,E_{S}\,\,$ are two
times higher then the corresponding $\,\,NN\,\,$ and
$\,\,pA\,\,$ values,a fact established from recent NA35
data. Therefore new ideas are required to modify hadronic
models in order to reproduce strange particle production
at high energies.

 \subsubsection{Multiplicity ratios}

 A different approach of finding signals for QGP is simply to
 compare nucleus-nucleus (AA) or hadron-nucleus(hA) data with
 nucleon-nucleon(NN) data.

 However,in particular this assumption is certainly not correct,
 since at high energies AA scattering is not simply a superposition
 of NN collisions.The time scale between two NN interactions is too
 small for hadronization and therefore some intermediate object is
 involved.

 Simple theoretical estimates are very unrealiable .Comparing AA
 with NN data or pA and NN data does not help,because new results
 do not necessarily mean QGP.Calculation based on QCD are not
 feasible,because we are in domaine where PQCD does not apply.
 Soft interactions involving small transverse momenta transfer
 should be taken into account.

 This ratios for multistrange particles were analised in terms
 of thermal models\cite{kn:raf94},\cite{kn:let94},\\
\cite{kn:heinz94},
 \cite{kn:soll93}.
 A constraint between thermal fireball parameters arises from the
 requirement that the balance of strangeness in a fireball is
 nearly zero.The impact of this constraint on (multi-)strange
 (anti-)baryon multiplicities comparing hadron gas and QGP has
 been analised.The data are compatible with the
 QGP hypothesis and appear to be inconsistent with the picture of
 an equilibrated hadron gas fireball \cite{kn:let94},\cite{kn:raf94}.

 However the two scenarios (slow and rapid)for the expansion of
 a QGP were recently studied in detail .Production ratios
 $\Lambda/\bar{\Lambda}$ and $\Xi/\bar{\Xi}$ for hadron gas at
 $T=200 MeV$ and plasma break-up have been compared to experimental
 data at baryonic chemical potential $\mu_{B}=300 MeV$.Present data
 on strange particle production cannot provide a distinction
 between the two scenarios \cite{kn:cley92}.

 We calculate the ratios of the mean strange particles
 multiplicities to the mean multiplicities of negative
 hadrons and we show in Table 6 a comparison with
 experimental NA35 data \cite{kn:na3594} for the same
 interactions like in Table 3.The agreement cannot be
 obtained without factor of two in neutral strange
 multiplicities . We note that negative hadrons as well as
 negative pions are well described even for very complex
$\,\,AA\,\,$ interactions.

 In Tabel 7 we give the multiplicity ratios for negative hadrons
 as well as for neutral strange particle for pA/pp ,AA/pp and AA/pA
  Resuts from HIJING
 ($HIJ.01$) and $DPM$ models are compared to NA5 \cite{kn:10na35}
 and NA35 \cite{kn:1na35} experimental data.The calculations on
 the models $HIJ.01$ and $DPM$ reproduce the relative increase of
 the $K_{s}^{0}$ ratios in central $SS$ collisions,but cannot
 describe the data concerning $\Lambda$ and $\bar{\Lambda}$
 generation.

 Comparing experimental data with theoretical prediction of the
 HIJING model we see that these ratios cannot be described
 witouth factor of two for $\,\,AA/pp\,\,$ and
 $\,\,AA/pA\,\,$.

 Data for charge kaon production have already been published by
 HELIOS Collaboration for $p+W$ and $S+W$ at $200\,\, GeV/N$ in the
 limited target rapidity region $y=1.0-1.5$ and by NA35 Collab.
 for $p+S,p+Au$and $S+S,S+Au$ collisions at $200\,\, GeV/N$ in a
 wider rapidity interval. Both experiments have shown an increase
 of the ratio $\frac{K^{+}}{\pi^{+}}$ as compared with the
 elementary interaction,while no significant increase was
 observed for $\frac{K^{-}}{\pi^{-}}$ ratio.

 The experimental values \cite{kn:5hel},\cite{kn:6hel} are compared
 with HIJING results in Tabel 8a and Tabel 8b.

 In Table 9 the results on average multiplicities of $K^{+}$ and
 $K^{-}$ and multiplicities ratios in $S+S$ interactions extrapolated
 to the full phase space are compared with model predictions :
 HIJING (HIJ.01),VENUS and FRITIOF. The experimental data are taken
 from reference \cite{kn:6na35}.
 All models underestimate the $\frac{<K^{+}>}{ <\pi^{+}>} $ ratio ,but
 reproduce well the $\frac{<K^{-}}{ <\pi^{-}>}$ ratio.
 The models seems to overpredict the average $<K^{-}>$ multiplicities
 in nucleus-nucleus collisions.The secondary collisions induced by
 the pion and kaon mesons in nuclei may influence the ratios
 $\frac{<K>}{<\pi>}$ \cite{kn:liu} .

 Experimental values for multistrange baryon ratios in nucleus
 nucleus interactions were recently reported \cite{kn:8wa85},
 \cite{kn:5na36},
\cite{kn:pro6}.We check if the model could describe
 these ratios ,but our calculations are for full phase space and
 experimental data are reported in limited rapidity range.
 We see from Table 10 that only ratios
 $\frac{\bar{\Lambda}}{\Lambda}$ are well described by the
 model.The model overestimate the productions of
 $\bar{\Xi^{-}}$ and $\bar{\Omega^{-}}$  .

  The ratios are well described for $\,\,\bar{\Lambda}/{\Lambda}\,\,$
 but the ratios for multistrange particles and their antiparticle
 could not be predicted by the model.A detailed analysis in
  limited rapidity range is required in order to draw some
  definite conclusions (calculations are now in progress).
  We remark that the ratios for multistrange particle production
  are not predicted by this version of HIJING model,nor by more
  sophisticated models like QGSM , RQMD ,DPM,SPM.

 \subsection{Some  distributions for  neutral strange
 particles in pA and AA interactions}

 A qualitative prediction for rapidity and transverse
 kinetic energy are performed at $\,\,200 \,\,GeV/Nucleon\,\,$
in $\,\,p-S\,\,,S-S,S-Ag\,\,$ and $\,\,S-Au\,\,$ collisions
 and are represented in comparison with experimental data
 from T.Alner et al.,\cite{kn:na3594}.

In Fig.3a,b,c,d we give normalized rapidity distributions for
$\,\,\Lambda\,\,$ produced in interactions quoted above :
Fig.3a for $\,\,p-S\,\,$; Fig.3b for $\,\,S-S\,\,$,
Fig.3c for $\,\,S-Ag\,\,$;and Fig.3d for $\,\,S-Au\,\,$.
In Fig.4a,b,c,d  and in Fig.5a,b,c,d we give the
rapidity distributions for $\,\,\bar{\Lambda}\,\,$
and $\,\,K_{s}^{0}\,\,$ respectively.

In Fig.6a,b,c,d ;Fig.7a,b,c,d and in Fig.8a,b,c,d
normalized distributions of transverse kinetic energy defined as
$T_{kin}=m_{T}-m_{0}$,are represented for the same interactions.
Theoretical HIJING values (HIJ.01) are multiplied by a factor
of two (HIJ.01*2) shown in the figures.The total number of events
generated in full phase space  are the same as
for the results from Table 3(see section 3.1).

 Transverse momentum distributions for $\,\,\Lambda\,\,$,
  $\,\,\bar{\Lambda}\,\,$ and  $\,\,K_{s}^{0}\,\,$
 particles produced in
 $\,\,p+S\,\,$ interactions are given in Figures 9a,b,c.
 The ratio of $\,\,\bar{\Lambda}\,\,$ to $\,\,\Lambda\,\,$
 rapidity distributions for $\,\,p+S\,\,$ collisions are
 done in Fig.10 .

 Analysing the figures (fig.3a,b,c,d -- fig.5a,b,c,d) we see
 that similarly to $p+S$ collisions,the rapidity spectra of
$K_{s}^{0}$ and $\bar{\Lambda}$ particles in $ \,\,S-S\,\,$
interactions have a sharp peak centered at mid rapidity
and their width are close to the ones for the proton
induced case.
 {} For $\Lambda$ particles the distribution is naturally
 symmetrical and its width is larger due to possible production of
 strange baryons in the target(projectile) fragmentation region.
 A detailed discussion of $S+S$ collision system was done also
 by Werner \cite{kn:wer7} and Amelin et al. \cite{kn:ame2},
\cite{kn:ame94}.The projectile and target mass asymmetry is
 reflected in the $\,\,\Lambda\,\,$ rapidity distributions
which for $\,\,p-S\,\,$ ,$\,\,S-Ag\,\,$and $\,\,S-Au\,\,$
interactions reach a maximum at about 1.0 which is
compatible with those remarked in experiment (about 1 to 1.5).
The $\,\,\bar{\Lambda}\,\,$ and $\,\,K_{s}^{0}\,\,$ rapidity
distributions are less sensitive to the initial asymmetry of the
system.

As follows from figures,the shape of rapidity for
kaons,lambdas and antilambdas are reproduced fairly well by
HIJING model ,but the model uderestimate the production of
$\Lambda$ in the central region of rapidity spectra,even
for proton-sulphur interaction.
New experimental data and also data on proton  production
in hadron-nucleus and nucleus-nucleus interactions are
required in order to draw a definite conclusions.

 For transverse kinetic energy distributions (taken into
 account the factor of two) we obtain
 a good agreement with the data except the low
 $m_{T}$ values and in some cases the values at the tails.
  The normalised  transverse momentum distributions
 for $p-S$ interactions are well described except some
discrepacies  seen for $\Lambda$ (see Fig.9a,b,c).

  The ratio of the $\bar{\Lambda}$ and $\Lambda$ yields
is suggested to reflect the net baryon density of the particle
source. In Fig.10 the ratio of $\bar{\Lambda}$ to $\Lambda$
rapidity distributions for $p+S$ is presented in comparison
with experimental data .The theoretical values are
higher than experimental ones.

 It should be stressed out that in the most string models
 the produced strings do not interact and decay independently.
 Introduction of some collective effects like string fusion
  \cite{kn:sor93a},\cite{kn:sor94a} or firecreacker model
 \cite{kn:tai94} can change the predictions essentially ,
especially for rapidity plateau heights and high multiplicity
distribution tails.

 Some analysis was  done recently by Sorge \cite{kn:sorge2},
 \cite{kn:sorge3}in the framework of the RQMD .A doubling of
 lambda production in the target fragmentation region for
  collision on heavy targets were found due to nuclear
 cascading of the produced mesons.Also in RQMD approach
 the resonances play an important role in creating strange
 quark pairs.Werner \cite{kn:wer7} has shown that rescattering
 should be considered at CERN-SPS energies with the effect of
 increasing neutral strange particles production especially
 in the central region.HIJING does not incorporate these
 ingredients.

No model seems to get a real description of
absolute values for neutral strange particle production
without additional mechanisms.

 Since recently \cite{kn:8wa85} the data on multiple strange particle
 production ($\Xi^{-},\bar{\Xi^{-}}$) was reported ,in figures 11a-b we
 give rapidity distribution for full phase space
 $\Xi^{-}$ (fig.11a) and $\bar{\Xi^{-}})$
 (fig.11b)for $S+S$ interaction . We can see from these figures
 that the ratio of
 multiplicities $<\bar{\Xi^{-}}>\over <\Xi^{-}>$ has strogly dependence
 on rapidity bins. Improvement of the statistics and considerations
 of limited rapidity experimental intervals ,
  are needed in order to due some more toward
 comparison with experimental data.

 For giving an idea about of the abundances of particles produced in
 the near planed experiments at CERN we have studied $\,\,Pb+Pb\,\,$
 interaction at $170\,\, GeV/N$ and in addition to multiplicities given
 in Tabel 3 and Tabel 5, normalised  rapidity distributions are
 given in figures 12a--b for ($\Lambda,\bar{\Lambda}$).
 The number of generated events was$\,\, 10^{3}\,\,$
and the results are represented for full phase space.

 The dependences   rapidity - transvers momentum ,which
 define  theoretical values for acceptances
  as well as their
 bidimensional projections are also  given in figures 13 a-b for
 $\Lambda$ , in figures 14 a-b for $\bar{\Lambda}$
 and  for all negatives charges in
 $\,\,S-S\,\,$,$\,\,S-W\,\,$ and $\,\,Pb-Pb\,\,$ interactions
 in figures 15a,b;16a,b and 17a,b(respectively) and for
  all positives charges in $Pb-Pb$ interactions in
 figures 18a-b.

 We hope that from such kind of correlations we will get
more inside the dynamics  of the systems and we
will find some limits of the model concerning
 neutral strange particles production.

\section{Conclusions}

In this report ,we give a systematic study of strange particle
production in pp,pA and AA collisions mainly at SPS CERN-energies,
using HIJING MONTE CARLO model developed for high energy collisions.
We analyse data from CERN experiments (NA35,NA36,\\
WA85) and we give
a detailed comparison for mean multiplicities of neutral strange
particle for pp,pS,pAg,pW,pAu,SS,SW and PbPb interactions. One
should note that HIJING MODEL includes neither hipotheses of QGP.
Strange particle abundances are quite well
described in pp and pA collisions
 and are underpredicted by a factor of two
  in nucleus-nucleus interactions ,by
this version (theoretical values for HIJ.01) .The ratios for
multistrange particles could not
be predicted by the model HIJ.01 .The abundances of $\Xi^{-}$ and
$\Omega^{-}$ are underpredicted and the abundances of
$\bar{\Xi^{-}}$ and $\bar{\Omega^{-}}$ are overpredicted .
More carreful analisis are necessary in limited rapidity
interval.

We present also some calculations for the ratios
 $\frac{<K^{+}>}{<\pi^{+}>}$ and
$\frac{<K^{-}>}{<\pi^{-}>}$ given by HIJING model compared with
other models predictions like VENUS and FRITIOF.All models
seems to overpredict $\frac{<K^{-}>}{<\pi^{-}>}$ ratios and
underpredict $\frac{<K^{+}>}{<\pi^{+}>}$ ratios .
Rescattering should be important even at SPS -energies.

A qualitative prediction for rapidity distributions of neutral
strange particles in $\,p+S\,$ , $\,S+S\,$,$\,S-Ag\,$,$\,S-Au\,$
$200 GeV$per nucleon are done compared also with
recent experimental data[140].
The models describes correctly the shapes of the rapidity in
central $\,\,A-A\,\,$ collisions for
$\,\,K_{s}^{0},\Lambda\,\,$ and $\,\,\bar{\Lambda}\,\,$
but absolute numbers for multiplicities
are underestimated by a factor of two
for $\,\,<\Lambda>,<\bar{\Lambda}>,<K_{s}^{0}>\,\,$ in symmetric
collisions and for $\,\,<\Lambda>,<\bar{\Lambda}>\,\,$ in
asymmetric ones.
Some discrepancies are seen for $\Lambda$ rapidity distribution
and transverse momentum distributions
even for $\,\,p+S\,\,$ interactions.The factor of two can not
be explained taken into account minijets and very central
events($b_{max}=0.0 fm$).

The theoretical model predictions HIJ.01  presented above
for central $\,\,S-W\,\,$,$\,\,S-Au\,\,$
and $\,\,Pb-Pb\,\,$ collisions give an idea for particle abundances in
experiments which are under the way or are planned in the
near future at CERN.

The HIJING model simulations are however still somewhat too low
compared to the data.Also the ratios for multistrange baryons
could not be described.
 New ideas are necessary  in order to improve the comparison of
 theoretical calculations with experimental data.The main goal is to
 increase the rate of neutral strange particle production by a factor
 of two in nucleus-nucleus interaction at $200\,\, GeV/nucleon$.
 We remark that other models (QGSM,VENUS ,DPM and PSM) include now
 some aditional hypothesis in order to improve the discrepancies
 (colective string-string interactions,double or multiple colour
 exchanges,string fusion ).

\section{Acknowledgements}

One of the authors (VTP) would like to thank to Professor Miklos
Gyulassy and Dr Xin-Nian Wang for providing the Monte Carlo
program HIJING and many helpful communications.

Also expresses his gratitude to Professor C.Voci for kind
invitation and acknowledge financial support from INFN -
Sezione di Padova,Italy where this work was  finished.

He is also indebted to Professor E.Quercigh for kind hospitality
in CERN(may 1994), where part of this paper has been done
 and for very useful discussions.

\newpage

\begin{center}
 {\large \bf Table Captions}\\
 \vskip 0.3cm
{\bf Table 1} \\
{\it Particle multiplicities for pp
  interaction at 200 GeV}
\vskip 0.3cm
\end{center}

\begin{center}
{\bf Table 2}  \\
{\it Particle composition of a typical event at} $\sqrt{s} =546\,\, GeV$
\end{center}
\vskip 0.3cm

\begin{center}
 {\bf Table 3.} \\
 {\it Average multiplicities for negative hadrons,
 neutral strange particle in  pp,pA and AA
 interactions predicted by HIJING  model are compared
 with experimental data.The values for $HIJ.01^{*}$ are
 theoretical values for $\Lambda,\bar{\Lambda}$ and$ K_{s}^{0}$
 multiplied by a factor two and the values for$ HIJ.01^{**}$
 are theoretical values for $\Lambda,\bar{\Lambda}$ multiplied
  by a factor two.}

 \end{center}
 \vskip 0.3cm

\begin{center}
 {\bf Table 4.} \\
 {\it Theoretical predictions of HIJING are compared
  with others recent theoretical calculations .}
\end{center}
 \vskip 0.3cm
\begin{center}
 {\bf Table 5.} \\
 {\it The mean multiplicities of negative pions and $E_{S}$
 ratios(see the text for definition) for nuclear collisions
 at 200 GeV per nucleon .The experimental data are from
 NA 35 Collaboration(reference[140]) and the NN data are from
 reference[32].}
 \end{center}
 \vskip 0.3cm
\begin{center}
 {\bf Table 6.} \\
 {\it  The ratios of the mean strange particle multiplicity
 to the mean multiplicity of negative hadrons.
 The values for $HIJ.01^{*}$ and $ HIJ.01^{**}$ have the same
 meaning as in Table 3.The experimental data are from T.Alner et al.
  [140]}
 \end{center}
 \vskip 0.3cm
 \begin{center}
 {\bf Table 7} \\
 {\it Multiplicity ratios for negative hadrons as well as
 for several identified strange particles and diferent reactions
 at $200 GeV/nucleon$;results from the HIJ.01 model are compared
 to DPM model[65], NA5[31]  and NA35[22] data}.
 \end{center}
 \vskip 0.3cm
 \begin{center}
{\bf Tabel 8a}  \\
 \end{center}
 \begin{center}
 $\frac{K}{\pi}$  {\it RATIOS}
 \end{center}
 \vskip 0.3cm
\begin{center}
 {\bf Tabel 8b} \\
 {\it Some ratios for pW and SW interactions at $200\,\, GeV/N$
  The experimental data are from reference [51]}
 \end{center}
 \vskip 0.3cm
\begin{center}
{\bf Table 9} \\
 {\it Average multiplicities of $K^{+}$ and $K^{-}$ and multiplicity
 ratios in $S+S$ interactions extrapolated to the full phase space.
 The experimental data are from Baechler et al.[27].
 The results for $p+p$ interaction [32] are also
  included for comparison.}
\vskip 0.3cm
\end{center}
\begin{center}
 {\bf Table 10}  \\
 \vskip 0.3cm
 {\it Multistrange multiplicity ratios at $200\,\, GeV/nucleon$.
 The experimental data are from NA35[22], NA36[127] and WA85[128,129]
 CERN experiments}.
 \end{center}

 \newpage

\begin{center}
{\bf Table 1} \\
\end{center}
\vskip 0.3cm
\begin{center}
\begin{tabular}{|c|c|c|c|}    \hline
{\bf pp} & {\bf Exp.data}  & {\bf HIJ.01} &${\bf HIJ.01^{(j)}}$ \\
\hline
 $<\pi^{-}>$  & $2.62\pm 0.06$ & $2.61$ & $2.65$ \\
 \hline
 $<\pi^{+}>$ & $3.22\pm 0.12$  & $3.18$ & $3.23$  \\
 \hline
 $<\pi^{0}>$  & $3.34 \pm 0.24$ & $3.27$ & $3.27$ \\
 \hline
 $<h^{-}>$  & $2.86 \pm 0.05$ & $2.99$ &$ 3.03$ \\
 \hline
 $<K^{+}>$  & $0.28 \pm 0.06$ & $0.32$ & $0.32$ \\
 \hline
 $<K^{-}>$  & $0.18 \pm 0.05$ & $0.24$& $0.25$  \\
 \hline
$<\Lambda + \Sigma^{0}>$ & $0.096 \pm 0.015$  & $0.16$& $0.165$ \\
 \hline
 $<\bar{\Lambda}+\bar{\Sigma^{0}}>$  & $0.013 \pm 0.01$  & $0.03$ &
  $0.037$  \\
 \hline
 $<K_{s}^{0}>$  & $0.17 \pm 0.01$ & $ 0.26$& $0.027$  \\
 \hline
 $<p>$  &           &$1.43$ & $1.45$ \\
 \hline
 $<\bar{p}>$  & $0.05 \pm 0.02$ & $0.11$& $0.12$  \\
 \hline
 $<\gamma>^{*}(dir)$ &     & $0.53$ & $0.55$ \\
 \hline
 $<\Xi^{-}>$  &      & $0.0015$ & $0.0019$ \\
 \hline
 $<\bar{\Xi^{-}}>$  &   & $0.0021$ & $0.0021$ \\
 \hline
 \end{tabular}
 \end{center}
 \vskip 0.3cm
 \newpage
\begin{center}
{\bf Table 2}  \\
\end{center}
\vskip 0.3cm
\begin{center}
\begin{tabular}{||c||c|c|c||}  \hline\hline
{\bf Particle type} & ${\bf <n>}$  & $ {\bf Exp.data}$
& $ {\bf HIJ.01^{(j)}}$  \\
\hline
\hline
{\bf All charged} &  $29.4 \pm 0.3$ & {\bf UA5} &  $28.2$  \\
\hline
$K^{0}+\bar{K^{0}}$ & $2.24 \pm 0.16$ & {\bf UA5} & $1.98$ \\
\hline
$K^{+}+K^{-}$&$ 2.24 \pm 0.16 $ & {\bf UA5}   & $2.06$ \\
\hline
$p+\bar{p}$ & $1.45 \pm 0.15$  & {\bf UA2} &  $1.55$ \\
\hline
$\Lambda+\bar{\Lambda}$ & $0.53 \pm 0.11$  & {\bf UA5} & $0.50$ \\
\hline$\Sigma^{+}+\Sigma^{-}+\bar{\Sigma^{+}}+\bar{\Sigma^{-}}$  &
 $0.27 \pm 0.06$  &        & $0.23$  \\
 \hline
 $\Xi^{-}$  &  $0.04 \pm 0.01$ & {\bf UA5} & $0.037$  \\
 \hline
 $\gamma$  & $33 \pm 3$  & {\bf UA5}  &  $29.02$  \\
 \hline
 $\pi^{+}+\pi^{-}$  &  $23.9 \pm 0.4$ &{\bf UA5} &   $23.29$ \\
 \hline
 $K_{s}^{0}$ &  $1.1 \pm 0.1$& {\bf UA5}  &   $0.99$  \\
 \hline
 $\pi^{0}$ & $11.0\pm 0.4$  &      &    $13.36$ \\
 \hline
 \hline
 \end{tabular}
 \end{center}
 \vskip 0.3cm
 \newpage
\begin{center}
 {\bf Table 3.} \\
\end{center}
 \vskip 0.3cm
\begin{center}
 \begin{tabular}{||c||c|c|c|c|c||}  \hline \hline
 ${\bf Reaction}$ &     & $<h^{-}>$ &  $<\Lambda>$ &
 $<\bar{\Lambda}>$ &$<K_{s}^{0}>$  \\
 \hline
  ${\bf p+p}$  & {\bf DATA} & $ 2.85 \pm 0.03$ &$ 0.096 \pm 0.015$ &
$ 0.013 \pm 0.005$ & $0.17 \pm 0.01$  \\
      &{\bf HIJ.01} & 2.99 & 0.16 & 0.030 & 0.26    \\
 \hline
 \hline
 ${\bf N+N}$ &{\bf DATA}& $3.22\pm0.06$ & $0.096\pm0.015$ &
   $0.013\pm0.005$& $0.20\pm0.03$   \\
        & $ {\bf HIJ.01}$ &  $3.29$ & $0.156$ & $0.033$ & $0.267$ \\
 \hline
 \hline
 ${\bf  p+S}$ &{\bf DATA} &$ 5.7 \pm 0.2$ &$ 0.28 \pm 0.03$ &
 $ 0.049 \pm 0.006$ & $0.38\pm0.05$ \\
 $'min. bias'  $ & ${\bf HIJ.01}$ &$4.83$  & $0.255$ & $0.046$ &
 $0.400 $ \\
 \hline
 \hline
 ${\bf  p+Ag}$& {\bf DATA} & $6.2\pm0.2$ & $0.37\pm0.06$ &
 $ 0.05 \pm 0.02 $ & $0.525 \pm 0.07 $ \\
 $'min. bias' $ & ${\bf HIJ.01}$ & $6.28$ & $0.34$ & $0.054$ &$ 0.505$\\
 \hline
 \hline
 ${\bf  p+Au}$& {\bf DATA} & $9.6 \pm 0.2$ &    &    &     \\
 $'central'$ & ${\bf HIJ.01}$ & $11.25$ & $0.67$ &$ 0.090$ & $0.88$  \\
 \hline
 \hline
          & {\bf DATA} & $95 \pm 5 $ & $9.4\pm1.0$ &$ 2.2\pm0.4$ &
  $10.5\pm1.7 $\\
 ${\bf  S+S}$  & ${\bf HIJ.01^{*}}$ & $88.8$ & $9.17$ & $1.73$ &
 $14.47$ \\
 $'central'$ & ${\bf HIJ.01^{**}}$ &$88.8$ &$9.17$ & $1.73$ &
 $ 7.23$  \\
        & ${\bf HIJ.01}$ &$ 88.8$ & $4.58$ & $0.86$ & $7.23$ \\
\hline
\hline
       & {\bf DATA} & $160\pm8$ & $15.2 \pm 1.2 $ &
        $ 2.6 \pm 0.3$ &$ 15.5 \pm 1.5 $ \\
${\bf  S+Ag}$ & ${\bf HIJ.01^{*}}$ & $ 164.35$ &$ 17.2$ &
  $2.97$ & $26.36$ \\
$'central'$ & ${\bf HIJ.01^{**}}$& $ 164.35$ &$ 17.2$ &
   $2.97$ & $13.20$ \\
      & ${\bf HIJ.01}$ &  $ 164.35$ & $8.61$ & $1.48$ &
   $13.20$ \\
\hline
\hline
       & {\bf DATA} &    &     &    &   \\
${\bf  S+Au}$ & ${\bf HIJ.01^{*}}$ & $213.2$ & $22.6$ & $3.62$ &
  $33.10$ \\
$'central'$ & ${\bf HIJ.01^{**}}$& $213.2$ & $22.6$ & $3.62$ &
 $16.55$ \\
      & ${\bf HIJ.01}$ & $213.2$ & $11.3$ & $1.81$ & $16.55$ \\
 \hline
 \hline
${\bf  S+W}$ & ${\bf HIJ.01^{**}}$ &$210.0$ & $21.28$ &$ 3.42$ &
 $16.05$ \\
$'central'$& ${\bf HIJ.01}$ & $210.0$ & $10.64$ &$ 1.71$ &
 $16.05$ \\
 \hline
 \hline
${\bf  Pb+Pb}$ & ${\bf HIJ.01^{*}}$ & $725.15$ & $72.88$ &
  $11.86$ & $109.7$ \\
$'central'$& ${\bf HIJ.01}$ &$725.15$ & $36.44$ &
  $5.93$ & $54.86$ \\
 \hline
 \hline
 \end{tabular}
 \end{center}
 \newpage

\begin{center}
 {\bf Table 4.} \\
\end{center}
\vskip 0.3cm
\begin{center}
\begin{tabular}{||c||c|c|c|c|c||}  \hline \hline
    &   &  &   &   &    \\
 ${ \bf  Reaction}$ &     &${ \bf  <h^{-}>}$ &
 $ { \bf  <\Lambda>}$ & ${ \bf <\bar{\Lambda}>}$
  &${ \bf  <K_{s}^{0}> }$ \\
 \hline
 \hline
       & {\bf DATA} & $ 2.85 \pm 0.03$ &$ 0.096 \pm 0.015$ &
      $ 0.013 \pm 0.005$ & $0.17 \pm 0.01$  \\
      &{\bf HIJ.01} &$ 2.99$ &$ 0.16$ &$ 0.030$ &$ 0.26  $  \\
     & {\bf RQMD} &$2.59$ & $0.11$ &    &$ 0.21$   \\
 ${\bf p+p}$  & {\bf QGSM} & $2.85$ & $0.15$ & $0.015$ & $0.21$  \\
     &$ {\bf DPM^{1}}$ & $3.52$ & $0.155$ & $0.024$ & $0.18$ \\
     &${\bf DPM^{2}}$ & $3.52$ & $0.155$ & $0.024$ & $0.18$ \\
     & ${\bf HIJ.01^{(j)}}$ & $3.03$ & $0.160$ &
       $0.037$ & $0.27$ \\
 \hline
 \hline
 ${\bf  p+S}$ &{\bf DATA} &$ 5.7 \pm 0.2$ &$ 0.28 \pm 0.03$ &
 $ 0.049 \pm 0.006$ & $0.38\pm0.05$ \\
 $'min. bias'  $ & ${\bf HIJ.01}$ &$4.83$  & $0.255$ & $0.046$ &
 $0.400 $ \\
      & {\bf RQMD} &    &     &     &     \\
& {\bf QGSM} & $5.87$ & $0.240$ & $0.023$ & $0.340$ \\
      &$ {\bf DPM^{1}}$ & $5.53$ & $0.300$ & $0.043$ &$ 0.360$ \\
      &${\bf DPM^{2}}$ & $5.54$ & $0.32$ & $0.060$ &$ 0.360$ \\
      & ${\bf HIJ.01^{(j)}}$ & $6.80$ & $0.37$ &
       $0.061$ & $0.57$ \\
 \hline
 \hline
       & {\bf DATA} & $95 \pm 5 $ & $9.4\pm1.0$ &$ 2.2\pm0.4$ &
  $10.5\pm1.7 $\\
 ${\bf  S+S}$  & ${\bf HIJ.01^{*}}$ & $88.8$ & $9.17$ & $1.73$ &
 $14.47$ \\
 $'central'$ & ${\bf HIJ.01^{**}}$ &$88.8$ &$9.17$ & $1.73$ &
 $ 7.23$  \\
        & ${\bf HIJ.01}$ &$ 88.8$ & $4.58$ & $0.86$ & $7.23$ \\
        &${\bf RQMD}$ & $110.2$ & $7.76$ &   & $10.0$  \\
        &${\bf QGSM}$ & $120.0$ & $4.70$ & $0.35$ & $7.0$ \\
        &${\bf DPM^{1}}$ & $109.8$ & $6.83$ & $0.80$& $10.6$ \\
        &${\bf DPM^{2}}$ & $107.0$ & $7.18$ & $1.57$ & $10.24$ \\
     &${\bf HIJ.01^{(j)}}$ & $89.5$ & $9.52$ & $1.76$
  & $7.47$ \\
\hline
\hline
      & {\bf DATA} & $160\pm8$ & $15.2 \pm 1.2 $ &
        $ 2.6 \pm 0.3$ &$ 15.5 \pm 1.5 $ \\
${\bf  S+Ag}$ & ${\bf HIJ.01^{*}}$ & $ 164.35$ &$ 17.2$ &
  $2.97$ & $26.36$ \\
$'central'$ & ${\bf HIJ.01^{**}}$& $ 164.35$ &$ 17.2$ &
   $2.97$ & $13.20$ \\
      & ${\bf HIJ.01}$ &  $ 164.35$ & $8.61$ & $1.48$ &
   $13.20$ \\
   &${\bf RQMD}$ & $192.3$ & $13.4$ &    &  $18.30$ \\
   &${\bf DPM^{1}}$ & $195.0$ & $13.3$ & $1.45$ & $19.40$ \\
   &${\bf DPM^{2}}$ & $186.90$ & $14.06$ & $3.65$ & $15.73$ \\
\hline
\hline
 \end{tabular}
 \end{center}
\newpage

\begin{center}
 {\bf Table 5.} \\
\end{center}
 \vskip 0.3cm
\begin{center}
 \begin{tabular}{||c||c|c|c|c|c||}  \hline \hline
 ${\bf Reaction}$ &     &  $<\pi^{-}>$ &
 $<E_{S}>$ &$<E_{S}^{*}>$ & $<E_{S}^{**}>$ \\
 \hline
 \hline
  ${\bf p+p}$  & {\bf DATA} & $ 2.62 \pm 0.06$ &
    &   &     \\
      &{\bf HIJ.01} & $ 2.61$ & $0.153$ &  &   \\
 \hline
 \hline
 ${\bf N+N}$ &{\bf DATA}&  $3.06 \pm 0.08 $ & $0.100\pm 0.01$ &
     &    \\
        & $ {\bf HIJ.01}$ &$2.89$  & $0.140$ &   &     \\
 \hline
 \hline
 ${\bf  p+S}$ &{\bf DATA} &$5.26\pm0.13$ &$ 0.086 \pm 0.008$ &
      &      \\
 $'min. bias'  $ & ${\bf HIJ.01}$ &$4.3$  & $0.144$ &    &    \\
 \hline
 \hline
${\bf  p+Ag}$& {\bf DATA} & $6.4\pm0.11$ & $0.108\pm0.009$ &
    &     \\
 $'min. bias' $ & ${\bf HIJ.01}$ & $5.59$ & $0.141$ &    &    \\
 \hline
 \hline
 ${\bf  p+Au}$& {\bf DATA} & $9.3 \pm 0.2$ &$0.073\pm0.015$ &   &  \\
 $'central'$ & ${\bf HIJ.01}$ & $10.22$ & $0.136$ & &   \\
 \hline
 \hline
          & {\bf DATA} & $88\pm5 $ & $0.183\pm0.012$ &  &  \\
 ${\bf  S+S}$  & ${\bf HIJ.01^{*}}$ & $79.6$ &  & $0.280$ &  \\
 $'central'$ & ${\bf HIJ.01^{**}}$ &$79.6$ &  &  & $0.160$ \\
        & ${\bf HIJ.01}$ &$ 79.6$ & $0.140$ &  &    \\
\hline
\hline
       & {\bf DATA} & $149\pm8$ & $0.173\pm0.017$ &
           &     \\
${\bf  S+Ag}$ & ${\bf HIJ.01^{*}}$ &$147.8$   &  &$0.277$  &  \\
$'central'$ & ${\bf HIJ.01^{**}}$& $147.8$ & &   & $0.158$  \\
      & ${\bf HIJ.01}$ &  $ 147.8$ &$0.138$  &  &    \\
\hline
\hline
& {\bf DATA} &    &     &    &   \\
${\bf  S+Au}$ & ${\bf HIJ.01^{*}}$ & $192.6$ &   & $0.268$ &  \\
$'central'$ & ${\bf HIJ.01^{**}}$& $192.6$ &  &  & $0.154$  \\
      & ${\bf HIJ.01}$ & $192.6$ & $0.105$ &  &  \\
 \hline
 \hline
${\bf  S+W}$ & ${\bf HIJ.01^{**}}$ &$179.84$ &  & & $0.158$ \\
$'central'$& ${\bf HIJ.01}$ & $179.84$ & $0.139$ & &  \\
 \hline
 \hline
${\bf  Pb+Pb}$ & ${\bf HIJ.01^{*}}$ & $621.75$ &  &$0.273$&  \\
$'central'$& ${\bf HIJ.01}$ &$621.75$ & $0.137$ &  &   \\
 \hline
 \hline
 \end{tabular}
 \end{center}
 \newpage

\begin{center}
 {\bf Table 6.} \\
\end{center}
 \vskip 0.3cm
\begin{center}
 \begin{tabular}{||c||c|c|c|c|c||}  \hline \hline
 ${\bf Reaction}$ &     & $<h^{-}>$ &  $\frac{<\Lambda>}{<h^{-}>}$ &
 $\frac{<\bar{\Lambda}>}{<h^{-}>}$ &$\frac{<K_{s}^{0}>}{<h^{-}>}$  \\
 \hline
  ${\bf p+p}$  & {\bf DATA} & $ 2.85 \pm 0.03$ &$ 0.034\pm0.005$ &
$ 0.0046\pm0.0018$ & $0.060\pm0.004$  \\
      &{\bf HIJ.01} &$ 2.99$ &$0.0535$  & $0.010$ & $0.0869$    \\
 \hline
 \hline
 ${\bf N+N}$ &{\bf DATA}& $3.22\pm0.06$ & $0.030\pm0.005$ &
   $0.0040\pm0.0016$& $0.062\pm0.009$   \\
        & $ {\bf HIJ.01}$ &  $3.29$ & $0.0483$ & $0.0101$ & $0.0816$ \\
 \hline
 \hline
 ${\bf  p+S}$ &{\bf DATA} &$ 5.7 \pm 0.2$ &$ 0.049\pm0.006$ &
 $ 0.086\pm0.0011$ & $0.067\pm0.001$ \\
 $'min. bias'  $ & ${\bf HIJ.01}$ &$4.83$  & $0.053$ & $0.0095$ &
 $0.0820 $ \\
 \hline
 \hline
${\bf  p+Ag}$& {\bf DATA} & $6.2\pm0.2$ & $0.0597\pm0.008$ &
 $ 0.0081\pm0.001$ & $0.0847\pm0.005 $ \\
 $'min. bias' $ & ${\bf HIJ.01}$ & $6.28$ & $0.0537$ &
     $0.0085$ &$ 0.0804$\\
 \hline
 \hline
    ${\bf  p+Au}$& {\bf DATA} &$7.0\pm0.4$& $0.060\pm0.007$&
    $0.010 \pm0.004$ & $0.0614\pm0.005$ \\
 $'min. bias'  $ &   ${\bf HIJ.01}$ & $7.34$ & $0.0535$ &
       $0.0084$ &$ 0.0761$ \\
 \hline
 \hline
 ${\bf  p+Au}$& {\bf DATA} & $9.6 \pm 0.2$ &    &    &     \\
 $'central'$ & ${\bf HIJ.01}$ & $11.25$ & $0.060$ &$ 0.0080$ & $0.078$  \\
 \hline
 \hline
          & {\bf DATA} & $95 \pm 5 $ & $0.099\pm0.012$ &$0.023\pm0.004 $ &
  $0.110\pm0.019$\\
 ${\bf  S+S}$  & ${\bf HIJ.01^{*}}$ & $88.8$ & $0.1032$ & $0.0194$ &
 $0.163$ \\
 $'central'$ & ${\bf HIJ.01^{**}}$ &$88.8$ &$0.1032$ & $0.0194$ &
 $ 0.0814$  \\
& ${\bf HIJ.01}$ &$ 88.8$ & $0.0516$ & $0.0097$ & $0.0814$ \\
\hline
\hline
       & {\bf DATA} & $160\pm8$ & $0.095\pm0.009 $ &
        $ 0.016\pm0.002$ &$ 0.097\pm0.011 $ \\
${\bf  S+Ag}$ & ${\bf HIJ.01^{*}}$ & $ 164.35$ &$ 0.1048$ &
  $0.0181$ & $0.0802$ \\
$'central'$ & ${\bf HIJ.01}$ &  $ 164.35$ & $0.0524$ & $0.0090$ &
   $0.0802$ \\
\hline
\hline
${\bf  S+Au}$  & {\bf DATA} &    &     &    &   \\
$'central'$ & ${\bf HIJ.01^{*}}$ & $213.2$ & $0.106$ & $0.0169$ &
  $0.0776$ \\
 \hline
 \hline
${\bf  S+W}$ & ${\bf HIJ.01^{**}}$ &$210.0$ & $0.1013$ &$0.0163 $ &
 $0.0764$ \\
$'central'$&  &  &  & &  \\
 \hline
 \hline
${\bf  Pb+Pb}$ & ${\bf HIJ.01^{*}}$ & $725.15$ & $0.1005$ &
  $0.0163$ & $0.1512$ \\
$'central'$&  & &  &  &    \\
 \hline
 \hline
 \end{tabular}
 \end{center}
 \newpage

\begin{center}
 {\bf Table 7} \\
\end{center}
 \vskip 0.3cm
\begin{center}
 \begin{tabular}{||c||c|c|c|c||}
 \hline \hline
 {\bf Ratio} & $<h_{-}>$ &  $<K_{s}^{0}>$ &
 $<\Lambda>$ &  $<\bar{\Lambda}>$  \\
 \hline
 \hline
 {\bf pS/pp} &$ 1.72\pm 0.07$ & $ 1.6 \pm 0.2$ &
 $2.3 \pm 0.4$ & $ 2.2 \pm 0.7$    \\
 \hline
 {\bf HIJ.01} & $1.61$ &$ 1.54$ & $1.594$ & $1.533$ \\
 \hline
 {\bf DPM } & 1.72 & 1.91 & 1.75 & 1.46 \\
 \hline
 \hline
 {\bf pAr/pp} &$ 1.88 \pm 0.18$ &$ 1.4 \pm 0.4$ &
 $2.4 \pm 0.8$ &     \\
 \hline
 {\bf HIJ.01} &     &     &     &      \\
 \hline
 {\bf DPM}  & 1.90 & 2.14 & 1.90 & 1.19 \\
 \hline
 \hline
 {\bf pXe/pp} &$ 2.39 \pm 0.06$ &$ 2.1 \pm 0.5$&
 $4.6 \pm 1.3$ &      \\
 \hline
 {\bf HIJ.01} &    &     &     &     \\
 \hline
 {\bf DPM}  &  2.62 & 2.94 & 2.39 & 1.31 \\
 \hline
 \hline
 {\bf pAu/pp} &$ 3.37 \pm 0.08$ &$ 2.3 \pm 0.6$ &
 $4.7 \pm 1.2$ &$ 3.4 \pm 1.3$      \\
 \hline
 {\bf HIJ.01} &$2.455$    &$2.15$  &$2.456$  &$2.06$    \\
 \hline
 {\bf DPM} &  2.81 & 3.18  &  2.39 & 1.23 \\
 \hline
 \hline
 {\bf pW/pp} &   &   &    &   \\
 \hline
 {\bf HIJ.01} & 3.61 &  3.07 &  3.69 &  2.43 \\
 \hline
 \hline
${\bf SS^{cent}/pp}$ &$ 36 \pm 2$ & $ 63 \pm 18$ &
 $86 \pm 12$ & $115 \pm 47$     \\
 \hline
 ${\bf HIJ.01^{*}}$ & $29.70$  &$55.61$   &$57.29$   &$57.56$   \\
 \hline
 {\bf DPM}  & 36.00  &  44.4 &  30.7 &  21.4 \\
 \hline
 \hline
  ${\bf SS^{cent}/pS}$ &$ 21 \pm 2$ &$ 38 \pm 10$ &
 $37 \pm 6$ &$ 54 \pm 16$      \\
 \hline
$ {\bf HIJ.01^{*}}$ & $18.38$ & $36.16$ & $35.95$ & $37.54$  \\
 \hline
 {\bf DPM}  &  20.90  & 23.20 & 17.60 & 14.60   \\
 \hline
 \hline
 \end{tabular}
\end{center}
\newpage

\begin{center}
 {\bf Tabel 8a}  \\
 \end{center}
\vskip 0.3cm
 \begin{center}
 \begin{tabular}{||c||c|c|c|c||} \hline
\hline
{\bf Interaction} & $\% \frac{K^{+}}{\pi^{+}}(exp)$ & $\% HIJ.01$ &
 $\% \frac{K^{-}}{\pi^{-}}(exp)$ & $\% HIJ.01$  \\
 \hline
 {\bf pp} & $10.8\pm 0.9$ & 9.98 & $8.6\pm 0.8 $ & 9.40 \\
 \hline
 {\bf pW} & $14.1 \pm 0.8 $ & 10.41 & $3.7 \pm 0.4 $ & 7.14 \\
 \hline
 {\bf SW} & $17.6 \pm 1.5 $ & 10.23 & $5.0 \pm 0.6 $ & 7.52 \\
 \hline
 \hline
\end{tabular}
 \end{center}
 \vskip 0.3cm
 \begin{center}
 {\bf Tabel 8b} \\
\end{center}
 \vskip 0.3cm
 \begin{center}
 \begin{tabular}{||c||c|c||} \hline\hline
 {\bf Ratio} &  \% {\bf Exp.data}  & {\bf \% HIJ.01}\\
 \hline
 \hline
 $\frac{\left(\frac{K^{+}}{\pi^{+}}\right)_{SW}}
   {\left(\frac{K^{+}}{\pi^{+}}\right)_{pW}}$ &
 $1.24 \pm 0.09 $ &  1.01  \\
 \hline
 $\frac{\left(\frac{K^{-}}{\pi^{-}}\right)_{SW}}
 {\left(\frac{K^{-}}{\pi^{-}}\right)_{pW}}$ &
 $1.31 \pm 0.22 $ & 1.42  \\
 \hline
 $\left(\frac{K^{+}+K^{-}}{\pi^{+}+\pi^{-}}\right)_{SW}  $ &
 12  &   8.99  \\
 \hline
 $\left(\frac{K^{+}+K^{-}}{\pi^{+}+\pi^{-}}\right)_{pW}  $ &
      8.9  & 8.77    \\
 \hline
 \hline
 \end{tabular}
 \end{center}
 \newpage

\begin{center}
{\bf Table 9} \\
\end{center}
\begin{center}
 \begin{tabular}{||c||c|c|c|c|c||}  \hline \hline
 ${\bf S+S}$  & $<K^{+}>$ & $<K^{-}>$ & $\frac{<K^{+}>}{<\pi^{+}>}$ &
 $\frac{<K^{-}>}{<\pi^{-}>}$ & $\frac{<K^{+}>}{<K^{-}>}$  \\
 \hline
 \hline
 {\bf Exp.data} &$ 12.5 \pm 0.4$ &$ 6.9 \pm 0.4$ &$ 0.137 \pm 0.008$ &
 $0.076 \pm 0.005$ & $1.81 \pm 0.12$ \\
 \hline
 {\bf HIJ.01} &$ 8.43$ & $6.27$ &$ 0.106$ &$ 0.0788$ &$ 1.345 $\\
 \hline
 {\bf VENUS} &$ 10.57 $ &$ 7.48$ &$ 0.117$ & $0.083$ & $ 1.41$ \\
 \hline
 {\bf FRITIOF} & $9.35$ & $7.02$ & $0.102$ & $0.077$&$ 1.33$  \\
 \hline
 {\bf p+p} &$ 0.28 \pm 0.06$ &$ 0.18 \pm 0.05$ &$ 0.087 \pm 0.02$ &
 $0.069 \pm 0.02$ &$ 1.55 \pm 0.5$ \\
 \hline
 {\bf HIJ.01} & $0.323$ & $0.24$ &$0.101$ &$0.093$ &$1.335$ \\
 \hline
 \hline
 \end{tabular}
 \end{center}
 \newpage

\vskip 0.3cm
\begin{center}
 {\bf Table 10}  \\
 \vskip 0.3cm
\end{center}

 \begin{center}
 \begin{tabular}{||c||c|c|c|c|c|c||} \hline \hline
 {\bf Reaction} & $\frac{\bar{\Lambda}}{\Lambda}$ & HIJ.01 &
 $\frac{\Xi^{-}}{\Lambda}$ & HIJ.01 &
 $\frac{\bar{\Xi^{-}}}{\bar{\Lambda}}$ & HIJ.01 \\
 \hline
 \hline
 {\bf pp} &$ 0.14 \pm 0.005$ & $0.18$ &    & 0.011 &
 $0.06 \pm 0.02$ & 0.056   \\
 \hline
 {\bf pS} &$ 0.175 \pm 0.02$ & 0.18 &   & 0.007 &
       &  0.078   \\
 \hline
 {\bf pAg} & $0.135\pm 0.022$ & $0.159$ &     &
      &   &   \\
 \hline
 {\bf pW} &        &  0.165 &     & 0.01 &
 $0.27 \pm 0.06$ & 0.066  \\
 \hline
{\bf SS} &$ 0.18 \pm 0.06$ & 0.186 &     & 0.009 &
        & 0.072   \\
 \hline
 {\bf SAg} &$0.171\pm 0.05$ & $0.172$     &       &
      &   &   \\
  \hline
  {\bf SW WA85} &$ 0.20 \pm  0.01$ & 0.163 &$ 0.09 \pm 0.01$ & 0.008 &
  $0.20 \pm 0.03$ & 0.070  \\
  \hline
  {\bf SW NA36} & $0.207 \pm 0.012$ & 0.163 &  $0.066 \pm 0.013$& 0.008 &
            & 0.070  \\
  \hline
  {\bf PbPb} &   & 0.163 &    & 0.006 &     &  0.067   \\
  \hline
  \hline
  \end{tabular}
  \end{center}

  \newpage
  {\large \bf Figure Captions}\\
  \vskip 0.3cm
  {\bf Fig.1\\}
 \vskip 0.3cm
  Rapidity distributions (Fig.1a) and transverse momentum
  distributions(Fig.1b) for $\Lambda$ particles produced in
  $\,\,pp\,\,$ interactions at $200\,\, GeV$.
  Experimental data are taken from Jaeger et al.[107].
  \vskip 0.3cm
   {\bf Fig.2}\\
 \vskip 0.3cm
   Rapidity distributions (Fig.2a)and transverse momentum
 distributions(Fig.2b) for $K_{s}^{0}$ particles produced in
 $\,\,pp\,\,$ interactions at $200\,\, GeV$.
 Experimental data are taken from Jaeger et al.[107].
 \vskip 0.3cm
  {\bf Fig.3}\\
 \vskip 0.3cm
 Rapidity distributions for $\,\,\Lambda\,\,$ particles produced in
 minimum bias $\,\,p-S\,\,$ (a) interactions and central
 $\,\,S-S\,\,$(b),$\,\,S-Ag\,\,$(c) and $\,\,S-Au\,\,$ (d)
 collisions at $200\,\,GeV$ per nucleon.The experimental data
 (full circles) are taken from T.Alner et al.,[140].
 The dashed histograms are theoretical
 values$ HIJ.01$  and the solid histograms are theoretical
 predictions $HIJ.01$ multiplied by a factor of two $ HIJ.01*2$.
  The open circles show the distributions
for $\,\,S-S\,\,$ collisions reflected at $y_{cm}=3.0$
 \vskip 0.3cm
{\bf Fig.4}\\
\vskip 0.3cm
 Rapidity distributions for $\,\,\bar{\Lambda}\,\,$
 particles produced in minimum bias $\,\,p-S\,\,$ (a) interactions
 and central $\,\,S-S\,\,$(b),$\,\,S-Ag\,\,$(c)
and $\,\,S-Au\,\,$ (d) collisions at $200\,\,GeV$ per nucleon.
 The experimental data (full circles) are taken from
  T.Alner et al.,[140].The dashed histograms are theoretical
 values$ HIJ.01$  and the solid histograms are theoretical
 predictions $HIJ.01$ multiplied by a factor of two $ HIJ.01*2$.
 The open circles show the distributions
for $\,\,S-S\,\,$ collisions reflected at $y_{cm}=3.0$
 \vskip 0.3cm
 {\bf Fig.5}\\
 \vskip 0.3cm
 Rapidity distributions for $\,\, K_{s}^{0}\,\,$ particles
 produced in minimum bias $\,\,p-S\,\,$ (a) interactions
 and central $\,\,S-S\,\,$(b),$\,\,S-Ag\,\,$(c)
and $\,\,S-Au\,\,$ (d) collisions at $200\,\,GeV$ per nucleon.
 The experimental data (full circles) are taken from
 T.Alner et al.,[140].The dashed histograms are theoretical
 values$ HIJ.01$  and the solid histograms are theoretical
 predictions $HIJ.01$ multiplied by a factor of two $ HIJ.01*2$.
The open circles show the distributions
 for $\,\,S-S\,\,$ collisions reflected at $y_{cm}=3.0$
 \vskip 0.3cm
 {\bf Fig.6}\\
 \vskip 0.3cm
 Transverse kinetic energy distributions for $\,\,\Lambda\,\,$
 particles produced in minimum bias $\,\,p-S\,\,$ (a) interactions
and central $\,\,S-S\,\,$(b) $\,\,S-Ag\,\,$(c)
 and $\,\,S-Au\,\,$ (d) collisions at $200\,\,GeV$ per nucleon.
 The experimental data (full circles) are taken from
T.Alner et al.,[140].
 The solid histograms are theoretical HIJING values ($HIJ.01$)
 or $HIJ.01*2$ shown in the figures.The vertical scale is given
 in $\,\,GeV^{-2} \,\,$.
 \vskip 0.3cm
 {\bf Fig.7}\\
 \vskip 0.3cm
 Transverse kinetic energy distributions for $\,\,\bar{\Lambda}\,\,$
 particles produced in minimum bias $\,\,p-S\,\,$ (a) interactions
 and central $\,\,S-S\,\,$(b) $\,\,S-Ag\,\,$(c)
and $\,\,S-Au\,\,$ (d) collisions at $200\,\,GeV$ per nucleon.
The experimental data (full circles) are taken from
T.Alner et al.,[140].
 The solid histograms are theoretical HIJING values ($HIJ.01$)
or $HIJ.01*2$ shown in the figures.The vertical scale is given
in $\,\,GeV^{-2} \,\,$.
 \vskip 0.3cm
 {\bf Fig.8}\\
 \vskip 0.3cm
 Transverse kinetic energy distributions for $\,\,K_{s}^{0}\,\,$
 particles produced in minimum bias $\,\,p-S\,\,$ (a) interactions
and central $\,\,S-S\,\,$(b) $\,\,S-Ag\,\,$(c)
and $\,\,S-Au\,\,$ (d) collisions at $200\,\,GeV$ per nucleon.
The experimental data (full circles) are taken from
 T.Alner et al.,[140].
The solid histograms are theoretical HIJING values ($HIJ.01$)
or $HIJ.01*2$ shown in the figures.The vertical scale is given
in $\,\,GeV^{-2} \,\,$.
\vskip 0.3cm
{\bf Fig.9}\\
\vskip 0.3cm
 Transverse momentum distributions for $\,\,\Lambda\,\,$ (a),
 $\,\,\bar{\Lambda}\,\,$ (b) and $\,\,K_{s}^{0}\,\,$ (c)
 particles produced in minimum bias $\,\,p-S\,\,$ interactions
at $200\,\,GeV$ per nucleon.The solid histograms are theoretical
HIJING values ($HIJ.01$).The vertical scale is given in
 $\,\,(GeV/c)^{-1}\,\,$.
 \vskip 0.3cm
 {\bf Fig.10}\\
 \vskip 0.3cm
 The ratio of $\,\,\bar{\Lambda}\,\,$  to $\,\,\Lambda\,\,$ rapidity
 distributions for $p-S$ interactions at $200\,\,GeV$ per nucleon.
 The corresponding ratio calculated using HIJING model is shown
by solid histogram. The experimental data [140] are represented by
 full circles.
 \vskip 0.3cm
 {\bf Fig.11}\\
 \vskip 0.3cm
 Rapidity distributions for $\,\,\Xi^{-}\,\,$ (a) and for
 $\,\,\bar{\Xi^{-}}\,\,$ (b) particles produced in
 $\,\,S--S\,\,$ interactions at $200\,\,GeV$ per nucleon.
 Histograms are HIJING results.

 \vskip 0.3cm
 {\bf Fig.12}\\
 \vskip 0.3cm
 Predicted rapidity distributions for $\,\,\Lambda\,\,$ (a)
 and for $\,\,\bar{\Lambda}\,\,$ (b) particle produced
in  central $\,\,Pb-Pb\,\,$ collisions at $\,\,170\,\,AGeV\,\,$
 \vskip 0.3cm
 {\bf Fig.13}\\
 \vskip 0.3cm
 Theoretical values for acceptances  are
 given (a) for $\,\,\Lambda\,\,$ particles produced in
 $\,\,Pb-Pb\,\,$ collisions at $\,\,170\,\,AGeV\,\,$.
 A bidimensional plot (rapidity y - transverse momentum
 $p_{T}$) (b) for the same interaction.
 \vskip 0.3cm
 {\bf Fig.14}\\
 \vskip 0.3cm
 Theoretical values for acceptances are
 given (a) for $\,\,\bar{\Lambda}\,\,$ particles produced
 $\,\,Pb-Pb\,\,$ collisions at $\,\,170\,\,AGeV\,\,$.
 A bidimensional plot (rapidity y - transverse momentum
 $p_{T}$) (b) for the same interaction.
 \vskip 0.3cm
 {\bf Fig.15}\\
 \vskip 0.3cm
 Theoretical values for acceptances are
 given (a) for all negatives charges produced in
 $\,\,S-S\,\,$ central collisions at $\,\,200\,\,AGeV\,\,$.
  A bidimensional plot (rapidity y - transverse momentum
 $p_{T}$) (b) for the same interaction.
 \vskip 0.3cm
 {\bf Fig.16}\\
 \vskip 0.3cm
 Theoretical values for acceptances are
given (a) for all negatives charges produced in
 $\,\,S-W\,\,$ central collisions at $\,\,200\,\,AGeV\,\,$.
 A bidimensional plot (rapidity y - transverse momentum
 $p_{T}$) (b) for the same interaction.
\vskip 0.3cm
 {\bf Fig.17}\\
 \vskip 0.3cm
 Theoretical values for acceptances are
 given  (a) for all negatives charges produced in
 $\,\,Pb-Pb\,\,$ central collisions at $\,\,170\,\,AGeV\,\,$.
 A bidimensional plot (rapidity y - transverse momentum
 $p_{T}$) (b) for the same interaction.
\vskip 0.3cm
 {\bf Fig.18}\\
 \vskip 0.3cm
 Theoretical values for acceptances are
 given  (a) for all positives charges produced in
 $\,\,Pb-Pb\,\,$ central collisions at $\,\,170\,\,AGeV\,\,$.
 A bidimensional plot (rapidity y - transverse momentum
 $p_{T}$) (b) for the same interaction.

  \end{document}